%% file: main.tex
\documentclass[conference]{IEEEtran}
\usepackage{epsfig}
\usepackage{endnotes}
\usepackage{makecell}
\usepackage{booktabs}
\usepackage{multirow}
\usepackage{array}
\usepackage{adjustbox}
\usepackage{tabularx}
\usepackage[framemethod=TikZ]{mdframed}
\usepackage{paralist}
\usepackage[most]{tcolorbox}
\usepackage{caption}
\usepackage{url}
\usepackage{breqn}
\usepackage{amsmath}
\usepackage{xspace}
\newtheorem{protocol}{Protocol}

\newcommand{\system}{{\ensuremath{\sf{KESIC}}}\xspace}
\newcommand{\systemtext}{{\underline{K}erberos \underline{E}xtensions for \underline{S}mart, \underline{I}oT and \underline{C}PS Devices\xspace}}

\newcommand{\idx}[1]{$ID_{#1}$\xspace}
\newcommand{\adc}{$AD_{c}$\xspace}
\newcommand{\iotserver}{ISV\xspace}
\newcommand{\pcd}{$Dev_{pc}$\xspace}
\newcommand{\gd}{$Dev_{g}$\xspace}
\newcommand{\plaingd}{Dev_{g}}
\newcommand{\plainpc}{Dev_{pc}}

\newcommand{\tsxy}[2]{$TS_{#1 \rightarrow #2}$\xspace}
\newcommand{\lfn}[1]{$L_{#1}$\xspace}
\newcommand{\reqxy}[2]{$Req_{#1 \rightarrow #2}$\xspace}
\newcommand{\resxy}[2]{$Res_{#1 \rightarrow #2}$\xspace}
\newcommand{\klxy}[2]{$K_{#1 \leftrightarrow #2}$\xspace}

\newcommand{\klxysync}[2]{$K_{#1 \leftrightarrow #2}$\xspace}
\newcommand{\kxy}[2]{$k_{#1 \leftrightarrow #2}$\xspace}
\newcommand{\kxyattest}{$k_{\plainpc \leftrightarrow \iotserver}$\xspace}
\newcommand{\tx}[1]{$T_{#1}$\xspace}
\newcommand{\axy}[2]{$A_{#1 \rightarrow #2}$\xspace}
\newcommand{\axyattest}[2]{$A_{#1 \rightarrow #2}^{attest}$\xspace}
\newcommand{\reqattest}{$Req_{\iotserver \rightarrow \plainpc}^{attest}$\xspace}
\newcommand{\resattest}{$Res_{\plainpc \rightarrow \iotserver}^{attest}$\xspace}
\newcommand{\csync}{$Co_{sync}$\xspace}
\newcommand{\cpcd}{$Co_{\plainpc}$\xspace}

\newcommand{\pidx}[1]{ID_{#1}\xspace}
\newcommand{\padc}{AD_{c}\xspace}
\newcommand{\ptsxy}[2]{TS_{#1 \rightarrow #2}\xspace}
\newcommand{\plfn}[1]{L_{#1}\xspace}
\newcommand{\preqxy}[2]{Req_{#1 \rightarrow #2}\xspace}
\newcommand{\presxy}[2]{Res_{#1 \rightarrow #2}\xspace}
\newcommand{\pklxy}[2]{K_{#1 \leftrightarrow #2}\xspace}
\newcommand{\pklxykey}[2]{K_{#1 \leftrightarrow #2}\xspace}
\newcommand{\pklxytkt}[2]{K_{#1 \leftrightarrow #2}\xspace}
\newcommand{\pklxysync}[2]{K_{#1 \leftrightarrow #2}\xspace}
\newcommand{\pkxy}[2]{k_{#1 \leftrightarrow #2}\xspace}
\newcommand{\pkxyattest}{k_{\plainpc \leftrightarrow \iotserver}\xspace}
\newcommand{\ptx}[1]{T_{#1}\xspace}
\newcommand{\paxy}[2]{A_{#1 \rightarrow #2}\xspace}
\newcommand{\paxyattest}[2]{A_{#1 \rightarrow #2}^{attest}\xspace}

\newcommand{\pcsync}{Co_{sync}\xspace}
\newcommand{\pcpcd}{Co_{\plainpc}\xspace}

\newcommand{\syncmanager}{SM\xspace}
\newcommand{\tktmanager}{TM\xspace}

\newcommand{\prv}{{\ensuremath{\sf{\mathcal Prv}}}\xspace}
\newcommand{\vrf}{{\ensuremath{\sf{\mathcal Vrf}}}\xspace}
\newcommand{\ra}{{\ensuremath{\sf{\mathcal RA}}}\xspace}
\newcommand{\chal}{{\ensuremath{\sf{\mathcal Chal}}}\xspace}
\newcommand{\sadv}{{\ensuremath{\sf{\mathcal Adv}}}\xspace}

\begin{document}
\title{\system: \systemtext}
\author{
{\rm Renascence Tarafder Prapty}\\
    University of California Irvine\\
    rprapty@uci.edu
    \and
    {\rm Sashidhar Jakkamsetti}\\
    Bosch Research\\
    sashidhar.jakkamsetti@us.bosch.com
    \and
    {\rm Gene Tsudik}\\
    University of California Irvine\\
    gts@ics.uci.edu
}
\maketitle

\begin{abstract}
\input{content/00-abstract}
\end{abstract}
\input{content/01-introduction}
\input{content/02-background}

\input{content/03-design_overview}

\input{content/04-protocols}

\input{content/05-implementation}

\input{content/06-evaluation}

\input{content/07-related_works}
\input{content/08-conclusion}
\bibliographystyle{unsrt}
\bibliography{main}
\end{document}

%% file: content/00-abstract.tex
Secure and efficient multi-user access mechanisms are increasingly important for the growing 
number of Internet of Things (IoT) devices being used today.

Kerberos is a well-known and time-tried security authentication and access control system for
distributed systems wherein many users securely access various distributed services. 
Traditionally, these services are software applications or devices, such as printers.
However, Kerberos is not directly suitable for IoT devices due to its relatively
heavy-weight protocols and the resource-constrained nature of the devices. 

This paper presents \system, a system that enables efficient and secure multi-user access for
IoT devices. \system aims to facilitate mutual authentication of IoT devices and users via Kerberos 
without modifying the latter's protocols. To facilitate that, \system includes a special 
Kerberized service, called IoT Server, that manages access to IoT devices.
\system presents two protocols for secure and comprehensive multi-user access system for 
two types of IoT devices: general and severely power constrained. In terms of performance, 
\system consumes $\approx~47$ times less memory,  and incurs $\approx~135$ times lower
run-time overhead than Kerberos.

%% file: content/01-introduction.tex
\section{Introduction}
Internet-of-Things (IoT) and Cyber-Physical Systems (CPS) encompass 
embedded devices with sensors, actuators, control units, and network connectivity. 
They are widely adopted in both private and public spaces. By 2030, the number of IoT devices is projected to surpass $29$ billion~\cite{statista-iot}. 

Except for personal devices, such as wearables, most IoT devices benefit from multi-user access. 
For instance, multiple family members can manage a smart home system comprised of smart TVs, 
voice assistants, and smart appliances. Similarly, in a shared office setting, many employees operate IoT devices, 
(e.g., smart projectors, meeting room schedulers, and interactive whiteboards). 
Maintenance crews or engineers often monitor IoT devices used for automation and process control in industrial 
settings, such as factories or warehouses. Multiple users being able to control devices provides convenience 
and facilitates coordination.

On another note, IoT devices usually have limited or no security features due to their resource constraint nature. This makes them vulnerable to different attacks, e.g., Mirai Botnet~\cite{Mirai'16}, Triton~\cite{Triton}. Therefore, it is crucial for users to be able to ensure the integrity of an IoT device before using it

With respect to multi-user support, one intuitive idea is to require each user to individually register with each device and establish a security context, e.g., by sharing a unique symmetric key. 
However, this requires linear amount of storage on the device, which is impractical due to resource constraints. 

To this end, some prior work introduced the notion of communication proxies~\cite{proxy_iot1}. 
Such proxies run a lightweight protocol between themselves and devices, acting as intermediaries 
when devices communicate with users/clients. One drawback is the presence of an additional intermediate 
hop for every user request. Another prior effort \cite{kerberos_iot} proposed extending Kerberos to 
IoT devices, which involves significant changes to the Kerberos protocol. Specifically, 
additional message exchanges are needed between the device and modified Kerberos servers to 
authenticate service tickets, thus significantly impacting device runtime performance and availability. 
Furthermore, aforementioned approaches do not inform the user about software integrity of IoT devices, i.e.,
users do not learn whether a given device is healthy or compromised. 

Motivated by aforementioned issues, this work revisits using Kerberos for IoT devices to 
enable multi-user support. We present \system: \systemtext\ -- a design that requires no changes to 
Kerberos and includes attestation of the device's software state as a built-in service. 
We prefer Kerberos to other multi-user authentication schemes (e.g. OAuth, or OpenID) because 
it is specifically designed for accessing both hardware and software resources within a network. 
The original concept of Kerberos closely aligns with the idea of a network of IoT devices,
such as a smart home/office or an factory, making it an efficient and easily adaptable protocol for IoT devices. 

Configuring and using devices directly as Kerberos application services is impractical 
due to the lack of required hardware features: many (especially lower-end) IoT devices do not have 
real-time and/or secure clocks necessary for verifying Kerberos tickets, or sufficient memory 
to host the entire Kerberos library. 
More generally, storage, memory, runtime, and network overhead incurred by Kerberos 
is also significant for targeted devices (see Section \ref{subsec: comparison}).

Thus, instead of modifying Kerberos, \system uses an IoT Server (\iotserver) as a Kerberos service, 
that manages access to all constituent IoT devices, while relying on Kerberos for user authentication.
After initial Kerberos login, a user requests a service ticket for \iotserver from the Kerberos Ticket
Granting Service (TGS). Next, the user asks \iotserver to grant specific (IoT) tickets to access the 
desired IoT device. As part of this process, the user can request an attestation report for the IoT device in order 
to verify the latter's software integrity before actually using it.

\system partitions IoT devices into two groups: general and power constrained (see Section 
\ref{subsubsec:type_of_devices}). This grouping is based on the power consumption, 
hardware resources, and activity time of the devices.

\system considers three types of communications in the proposed ecosystem: user $\leftrightarrow$ \iotserver, 
\iotserver $\leftrightarrow$ IoT device, and user $\leftrightarrow$ IoT device. 
\system includes a protocol for each device category, covering all three interactions, resulting in two protocols.
\system does not require real-time clocks on devices: at each boot/wake-up time, 
devices obtain current time from \iotserver. 
Afterward, devices equipped with a timer use it to emulate a local clock.
For devices that are more power constrained and have no timers, we use a nonce-based approach.

Also, in terms of cryptography, \system exclusively uses keyed hash (HMAC) operations in all of its protocols,
instead of encryption and decryption operations, which can be slower and incur higher storage and run-time memory
footprints. Based on our proof-of-concept implementation, \system incurs $47$ times lower memory overhead and 
$135$ times lower run-time overhead than Kerberos.

Expected contributions of this work are: 
\begin{compactitem}
    \item Design of two lightweight and secure protocols for\ multi-user access to IoT devices.
    \item Open-source implementation of \system~\cite{kesic-repo} which includes: (i) a prototype \iotserver 
    integrated with Kerberos as an application service, (ii) two prototypes for IoT devices based on ARM 
    Cortex-M33 that uses TrustZone-M for implementing an attestation RoT, and (iii) a client 
    application that 
    uses \iotserver to obtain IoT tickets, and then requests access to IoT devices.
\end{compactitem}

%% file: content/02-background.tex
\section{Background}
This section overviews Kerberos and Root-of-Trust concepts. 
It can be skipped with no loss of continuity.

\subsection{Kerberos} \label{sec: traditional kerberos}
Kerberos is an authentication, authorization, and access control (AAA) system for distributed systems
that originated at MIT Project Athina in mid-1980s. It allows users to authenticate themselves once 
via username/password
(via Single Sign-On aka SSO or login) for all services that they are allowed to use. Once a user logs in,
no further human interaction is required for the duration of the login session. All other protocols 
used to access services are transparent to the (human) user.
There are three types of entities in Kerberos:

\textbf{Client or Principal (C)}: This is usually represented by a software component
called \textit{kinit} which resides in the user's workstation. It manages Kerberos tickets for users.

\textbf{Service Provider or Application Server (V)}: This software manages resources or services, e.g., graphic software or printer, and grants access to them by handling Kerberos service tickets.

\begin{figure}
    \centering
    \includegraphics[width=0.5\textwidth]{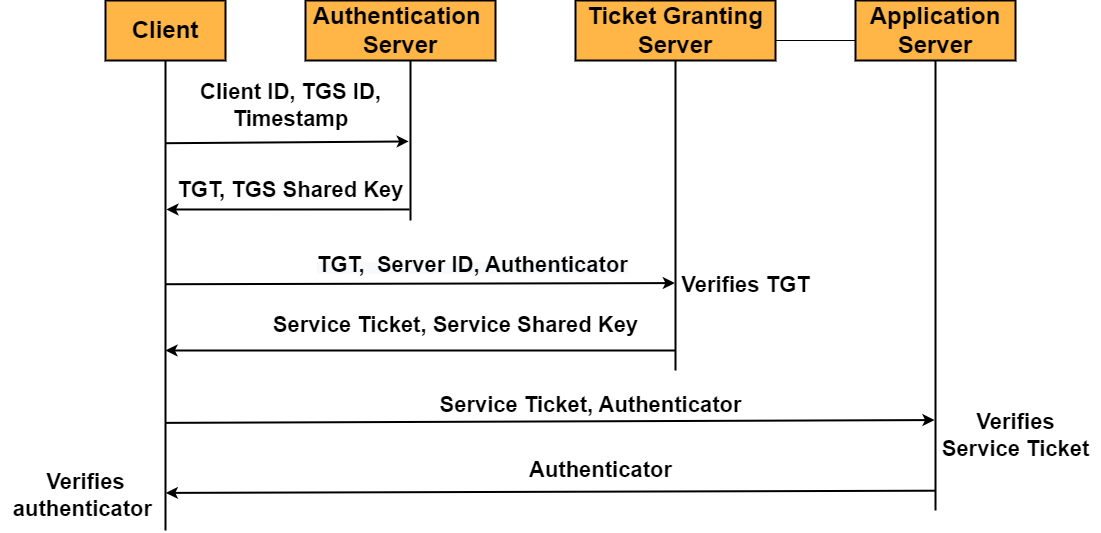}
    \caption{Kerberos Steps.}
    \vspace{-.4cm}
    \label{fig:kerberos common steps}
    
\end{figure}

\textbf{Key Distribution Center (KDC):} The central third party trusted by both clients and 
application servers. KDC maintains a database to store all user passwords and long-term keys for 
application servers. It also facilitates mutual authentication among clients and servers. 
KDC can also implement granular access control. It has two main components:
    \begin{compactitem}
        \item \textbf{Authentication Server (AS)} authenticates clients and issues Ticket Granting 
        Tickets (TGT).
        \item \textbf{Ticket Granting Server (TGS)} verifies TGTs and issues service ticket (\tx{V}) to access V.
    \end{compactitem}
Figures \ref{fig:kerberos common steps} and \ref{fig:protocol-traditional-kerberos} show an overview and 
description of Kerberos functionality, respectively, and Table \ref{table:notations} shows the notation.

\begin{figure}
  \captionsetup{justification=centering}
   \begin{mdframed}[linewidth=1pt]
    \begin{protocol}\label{protocol:protocol-traditional-kerberos}
    \footnotesize
    Kerberos Functionality:\\
    \textbf{C $\leftrightarrow$ AS Interaction}
    \begin{compactenum}
        \item C initiates a session, receives \idx{C} from user, builds \reqxy{C}{AS} and sends it to AS: $\preqxy{C}{AS} = \pidx{C} || \pidx{TGS} || \ptsxy{C}{AS}$ 
        \item Upon receiving \reqxy{C}{AS}, AS checks \idx{C} and retrieves the password related to C using the database, derives long-term key \klxy{C}{AS} from the password, and generates session key \kxy{C}{TGS}
        \item Then AS generates encrypted TGT, \tx{TGS}:\\
            $E(\pklxy{TGS}{AS}, [\pkxy{C}{TGS} || \pidx{C} || \padc || \pidx{TGS} || \plfn{2} ||$ \\ $\ptsxy{AS}{C}]) $
            
        \item Finally AS builds encrypted \resxy{AS}{C} and sends it to C: \\
            $\presxy{AS}{C} = E(\pklxy{C}{AS}, [\pkxy{C}{TGS} || \ptsxy{AS}{C} || \plfn{2} || \ptx{TGS}])$
        \item C receives \resxy{AS}{C}, receives password from user, and derives long term key \klxy{C}{AS} from password.
        \item \label{list: authentication through decryption} Then C decrypts \resxy{AS}{C} with \klxy{C}{AS}, retrieves and caches \kxy{C}{TGS}, \tx{TGS}.
    \end{compactenum}
    \textbf{C $\leftrightarrow$ TGS Interaction}
    \begin{compactenum}
        \item C computes an encrypted authenticator \axy{C}{TGS}: \\
            $\paxy{C}{TGS} = E(\pkxy{C}{TGS}, [\pidx{C} || \padc || \ptsxy{C}{TGS}])$
        \item Then C builds \reqxy{C}{TGS} and sends it to TGS: \\
            $\preqxy{C}{TGS} = \pidx{V} || \ptx{TGS} || \paxy{C}{TGS}$
        \item \label{list: verifying TGT} TGS receives \reqxy{C}{TGS} and decrypts \tx{TGS} with long term key \klxy{TGS}{AS}, verifies it by checking \lfn{2} and \idx{TGS}.
        \item \label{list: auth in tgs} Then TGS decrypts authenticator \axy{C}{TGS} with session key \kxy{C}{TGS} and verifies it by checking \idx{C}, \adc.
        \item After that TGS determines C's eligibility to access V, generates session key \kxy{C}{V}, and issues encrypted service ticket \tx{V}:
        \begin{equation*}
            E (\pklxy{V}{TGS}, [\pkxy{C}{V} || \pidx{C} || \padc || \pidx{V} || \ptsxy{TGS}{C} || \plfn{4}])
        \end{equation*}
        \item Finally, TGS builds encrypted \resxy{TGS}{C} and sends it to C: \\
                $\presxy{TGS}{C} = E(\pkxy{C}{TGS}, [\pkxy{C}{V} || \pidx{V} || \ptsxy{TGS}{C} || \ptx{V}])$
        \item \label{list: service ticket} Upon receiving \resxy{TGS}{C}, C decrypts it with session key \kxy{C}{TGS}, retrieves and caches \kxy{C}{V}, \tx{V}.
    \end{compactenum}
    \textbf{C $\leftrightarrow$ V Interaction}
    \begin{compactenum}
        \item C computes encrypted authenticator \axy{C}{V}: \\
        $\paxy{C}{V} = E(\pkxy{C}{V}, [\pidx{C}|| \padc || \ptsxy{C}{V}])$
        \item Then C builds \reqxy{C}{V} and sends it to V: \\
        $\preqxy{C}{V} = \ptx{V} || \paxy{C}{V}$
        \item \label{list: verifying service ticket} Upon receiving \reqxy{C}{V}, V decrypts \tx{V} with \klxy{V}{TGS}, verifies it by checking \lfn{4} and \idx{V}. 
        \item \label{list: auth in v} Then V decrypts authenticator \axy{C}{V} with session key \kxy{C}{V} and verifies it by checking \idx{C}, \adc.
        \item After that V computes an encrypted authenticator \axy{V}{C} and sends it to C for mutual authentication:\\ 
        \axy{V}{C} $=$ E(\kxy{C}{V}, [\tsxy{C}{V} + 1])
        \item \label{list: last step} C receives \axy{V}{C}, decrypts it with session key \kxy{C}{V} and verifies it by checking the decrypted value.
    \end{compactenum}
    \end{protocol}
  \end{mdframed}
  \vspace{-.2cm}
  \caption{Kerberos Protocol.}
  \label{fig:protocol-traditional-kerberos}
  \vspace{-.4cm}
\end{figure}

\textbf{Interaction between C \& AS}:
At the beginning of the Kerberos authentication process, C is authenticated by AS and provided a TGT (\tx{TGS}) for the next steps. AS does not explicitly authenticate C. 
Rather, successful decryption of \resxy{AS}{C} by C in step \ref{list: authentication through decryption} of C$\leftrightarrow$AS Interaction part
of Protocol \ref{protocol:protocol-traditional-kerberos} indicates that \klxy{C}{AS} is correct. Hence, the user 
provided the correct password and is authenticated.  Subsequent steps in the same session do not require
using the user password and \klxy{C}{AS}. Thus, the user needs to enter the password and authenticate
only once per session. Session duration is determined by the lifetime of \tx{TGS}. 

\textbf{Interaction between C \& TGS}: 
After being authenticated by AS and receiving TGT (\tx{TGS}), C can call TGS using the TGT. The purpose of this call to TGS is to request Kerberos tickets (\tx{V}) for different services.  
In step \ref{list: verifying TGT} of C$\leftrightarrow$TGS Interaction part
of Protocol \ref{protocol:protocol-traditional-kerberos}, TGS verifies \tx{TGS} 
by checking that \idx{TGS} is correct i.e. ticket is issued for the correct TGS, and ticket 
lifetime \lfn{2} is later than the current time, which implies the ticket is not expired. 
Afterward, TGS verifies authenticator \axy{C}{TGS} in step \ref{list: auth in tgs} by 
matching \idx{C} and \adc from \axy{C}{TGS} with corresponding values from \tx{TGS}. 
If both \axy{C}{TGS} and \tx{TGS} are valid, then C is authenticated to TGS. 
On the other hand, if either of these verifications fails, then C fails authentication, and TGS discards 
\reqxy{C}{TGS}. 

\textbf{Interaction between C \& V}:
When a user requires a service, C calls the application server V with the corresponding Kerberos service ticket \tx{V}.
In step \ref{list: verifying service ticket} of this interaction, V verifies \tx{V} by checking that \idx{V} is 
correct and \lfn{4} is later than the current time, i.e. ticket is not expired. Next, V 
verifies \axy{C}{V} in step \ref{list: auth in v} by matching \idx{C} and \adc from \axy{C}{V} with 
corresponding values from \tx{V}. 
If both \axy{C}{V} and \tx{V} are valid, then C is authenticated to V. On the other hand, 
if either of these verifications fails, then C fails authentication, and V discards \reqxy{C}{V}. 
\textit{\textbf{Checking \lfn{4} requires V to have a real-time clock}}. Because of this requirement, 
IoT devices without a real-time clock can not be directly configured as Kerberos application servers. 
Moreover, an expired \tx{V} can not be detected before it is decrypted, which would also cause 
unnecessary computational overhead for an IoT device. 

In step \ref{list: last step}, C verifies \axy{V}{C} by checking that the 
decrypted value is greater than \tsxy{C}{V} by 1. 
This verification authenticates V to C, because only genuine V should know \klxy{V}{TGS} and be able to decrypt \tx{V}.
It then lets V obtain \kxy{C}{V} and \tsxy{C}{V}, and compute \axy{V}{C}. 
After this step, C and V are mutually authenticated and C is granted access to V. 
Moreover, C and V share a secret key that can be used to secure further communication.
 
\subsection{Root-of-Trust (RoT)}
In \system, a RoT on an IoT device is needed for secure storage and secure computation. We need to ensure that shared long-term secret keys are not revealed and other authentication related metadata (such as synchronization value and counter) are not modified by potentially present malware. During attestation, we also need to securely compute HMAC of specified memory region without 
interference from any possibly present malware.

\system can be applied to three types of devices:
\begin{compactitem}
    \item Devices equipped with verified, secure, hybrid (SW/HW) RoTs, such as SANCUS 
    \cite{noorman2013sancus}, PISTIS \cite{pistis}, VRASED \cite{vrasedp,magpaper}, or RATA \cite{rata}.
    \item Off-the-shelf devices with hardware RoTs, such as ARM TrustZone\cite{trustzone}, 
    Intel SGX \cite{sgx}, or AMD SEV \cite{sev}.
    \item Legacy devices without any hardware RoT. In this case, there are two options: (i) rely on verified RoTs \cite{parsel} based on trustworthy microkernels \cite{seL4}, or (ii) however aspirational this might be, consider the OS to be trusted.     
\end{compactitem}  
It is to be noted that having hybrid or hardware RoT is not a prerequisite for using \system in IoT devices. 
Even though our proof of concept implementation uses TrustZone-M as RoT, \system is equally applicable to IoT devices where OS is trusted.

{\textbf{Remote Attestation (\ra)}}
is a security service that allows a trusted client (aka, verifier or \vrf) to 
measure software integrity on a remote device (aka, prover or \prv). 
\ra is a challenge-response protocol, usually realized as follows: 
\begin{compactitem}  
  \item \vrf sends an \ra request with a challenge (\chal) to \prv.
  \item \prv receives the request, computes an authenticated integrity check over its program 
  memory region and \chal, and returns the result to \vrf.
  \item \vrf checks the result and determines whether \prv is compromised.
\end{compactitem}

The integrity check is computed via either a Message Authentication Code (e.g., HMAC) or a 
digital signature (e.g., ECDSA) over \prv program memory. 
The former requires a long-term symmetric key shared between \prv and \vrf. For the latter, \prv must have 
a private key that corresponds to a public key known to \vrf. \ra architectures for 
low-end MCUs\cite{noorman2013sancus,vrasedp} use MACs whereas higher-end TEEs (such as  
Intel SGX and AMD SEV) use signatures. Both approaches require secure key storage on \prv.

%% file: content/03-design_overview.tex
\section{Design Overview} \label{sec:design}
As mentioned earlier, Kerberos in its regular incarnation, is unsuitable for low-end IoT devices for several reasons: 
(1) most IoT devices do not have real-time clocks and cannot verify timestamps on tickets, 
(2) storage and memory of IoT devices are limited and can not accommodate the Kerberos library 
(see Section \ref{subsec: comparison} 
for details), (3) since Kerberos tickets are encrypted, an expired ticket can not be detected until it is 
decrypted, which is time and resource-consuming, especially, for mission-critical devices. In fact, this could be 
abused as a means of DoS attacks.

Therefore, we opt to extend Kerberos -- without modifying it -- to support low-end IoT devices. Additionally,
since IoT devices are increasingly subject to malware attacks, we want to provide attestation of device software
to assure the user that the device is not compromised prior to its use.

\subsection{System Model}
\begin{figure}
    \centering
    \includegraphics[width=0.5\textwidth]{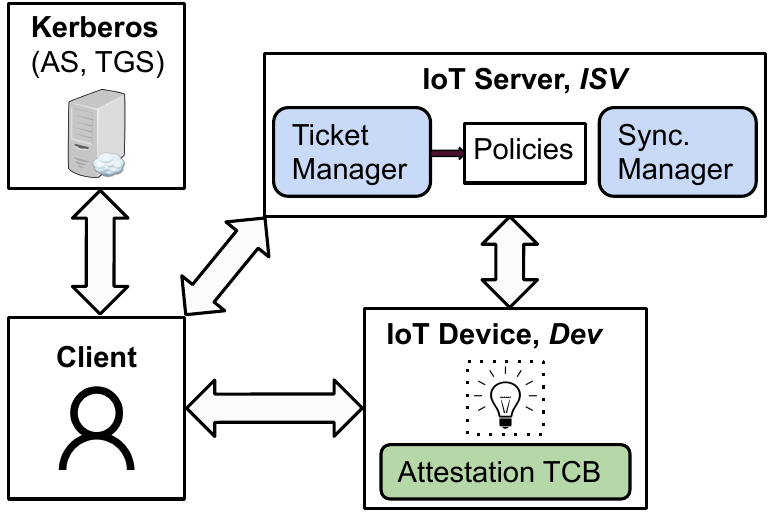}
    \caption{System Model.}
    \vspace{-.3cm}
    \label{fig: system model}
    \vspace{-.1cm}
\end{figure}

Figure \ref{fig: system model} overviews \system system model. As described before in Section \ref{sec: traditional kerberos}, Client (C) is a software component running in the human user's device (i.e. workstation or cellphone). We use the terms user and client interchangeably to refer to the same entity.

\subsubsection{IoT Server (\iotserver)} \label{subsubsec:ISRV}
We introduce a special Kerberos application service called IoT Server -- \iotserver. 
After initial log-in, C obtains a regular Kerberos ticket for \iotserver from TGS.
\iotserver is responsible for granting users access to IoT devices and managing authenticated communication 
between users and IoT devices. \iotserver has two main components:

\textbf{Ticket Manager (\tktmanager)} is responsible for authenticating users (clients) 
and issuing IoT tickets to access devices. Before granting an IoT ticket to a user, \tktmanager 
uses access control policies set up by the device owner to ensure that the user is permitted to 
access that device. These access control policies can be either static or dynamic. 
While important, access control policies are out of the scope of this paper and are not discussed further.
\system uses HMACs instead of encryption to generate tickets and authenticators. \tktmanager 
shares two long-term secret keys with each IoT device: one to generate IoT tickets for that device, 
and the other -- to generate the corresponding session key. 

\textbf{Synchronization Manager (\syncmanager)} is responsible for time synchronization between an 
IoT device and \iotserver, which determines the validity (i.e., freshness) of IoT tickets. 
\syncmanager maintains a distinct long-term key with each IoT device in order to secure the synchronization process.
Synchronization details vary depending on the type of the device; this is discussed in Section \ref{sec: main protocol}.
The Synchronization phase is crucial since the device cannot enter the service phase without 
successfully completing it. 

Indeed, \iotserver represents a single point of failure: if it is down, devices become non-operational.
This risk can be mitigated by deploying multiple \iotserver instances. 

\subsubsection{Device Types} \label{subsubsec:type_of_devices}
\system supports two types of IoT devices:

\textbf{General Devices (\boldmath{\gd})} are always awake; they have a direct power source or a long-lasting
battery. They can receive requests over the network at any time. Examples of such devices are Blink Security Camera~\cite{blink-security-camera}, Google Nest Thermostat~\cite{google-nest-thermostat}, and 
Lumiman Smart Bulb~\cite{lumimanbulb}.
 
\textbf{Power Constrained Devices (\boldmath{\pcd})} spend most of their time in a low-power sleep state 
due to stringent energy constraints. Periodically, they wake up, perform brief tasks, and return to sleep. 
Some can only receive network requests during active periods, while others are equipped with a separate network interface 
that briefly awakens the device in response to incoming requests. Examples of such devices are: 
ThermoPro TP357 Digital Hygrometer Indoor Thermometer~\cite{ThermoPro}, Netatmo Weather Station~\cite{Netatmo}, and
Nordic nRF9160 system-in-package (SiP)~\cite{nRF9160}.

There are two ways for clients to learn when a \pcd{} is awake: 
\begin{compactitem}
    \item when \pcd{} wakes up, it sends a notification to \iotserver. Then, \iotserver either broadcasts a notification to all clients or only grants tickets to \pcd that are awake.
    \item Client determines when a given \pcd{} should be awake based on prior information about \pcd{} sleep/awake schedule.
\end{compactitem}
Both \gd and \pcd devices are assumed to be equipped with either a hardware RoT or 
a trusted OS (in case of legacy devices). For the sake of simplicity, 
from now on, we refer to both of them as RoT.

\begin{table}
    \caption{Notation Summary.} \label{table:notations}
    \vspace{-1em}
    \footnotesize
    \linespread{0.8}
    \renewcommand{\arraystretch}{1.1}
    \begin{tabularx}{0.5\textwidth}{lX}
    \toprule
    Notation & Description    \\ 
    \midrule
    \idx{x} & Identity of entity x \\
    \adc & Identity of the Client Interface \\ 
    \gd & General Device \\
    \pcd & Power Constrained Device \\
    \tsxy{x}{y} & Timestamp sent from x to y \\ 
    \lfn{n} & Timestamp when the ticket will expire \\
    \reqxy{x}{y} & Request from x to y \\
    \resxy{x}{y} & Response from x to y \\
    \klxy{x}{y} & Long-term key between x and y \\
    \kxy{x}{y} & Session key between x and y \\
    \tx{x} & Ticket for entity x \\
    \axy{x}{y} & Authenticator sent from x to y \\
    \axyattest{x}{y} & Authenticator sent from \iotserver to \pcd and vice versa as part of attestation request/response \\
    \reqattest & Attestation Request from \iotserver to \pcd \\
    \resattest & Attestation Response from \iotserver to \pcd \\
    \cpcd & Counter maintained by \iotserver for \pcd{}, used to issue tickets \\
    \csync & Synchronized counter between \iotserver and \gd/\pcd, stores the number of synchronization requests \\
    \bottomrule
    \vspace*{-0.6cm}
    \end{tabularx}
\end{table}

\subsubsection{Protocol Overview}
\system has three run-time phases:

In \textbf{Ticket Issuing Phase}, C obtains a service ticket ($T_{\iotserver}$) for \iotserver after AS \& TGS log-in. 
Next, C obtains IoT tickets from \iotserver. IoT tickets issued for \gd and \pcd have different formats. 

In \textbf{Synchronization Phase}, which happens upon each boot, 
a device communicates with \iotserver to obtain the latter's current timestamp, which serves 
as the synchronization value for the current session. 
After that, a device uses a local timer (\gd) or a counter array (\pcd) to emulate a clock.

In \textbf{Service Phase}, an IoT device accepts IoT tickets and service requests from clients. 
\gd uses the synchronized local clock and \pcd uses its synchronized local counter array to 
determine ticket validity.

Section \ref{sec: main protocol} presents protocols covering each phase for both device classes.

\subsection{Adversary model} \label{sec: adversary model}
\system comprises multiple clients, application services, and IoT devices. 
It also includes trusted third parties: Authentication Server (AS), 
Ticket Granting Server (TGS), and IoT Server (\iotserver). 
We assume an adversary \sadv that can remotely attack an IoT device and compromise its software.
However, \sadv cannot attack any software and data inside the device RoT.
Furthermore, \sadv can compromise clients with the intent of:
\begin{compactitem}
    \item \textbf{Impersonate Clients} in order to circumvent access policy regarding a service/IoT device. 
    \item \textbf{Eavesdrop} on other clients’ message exchanges to obtain confidential information,
    such as passwords and keys.
    \item \textbf{Tamper} with other clients’ message exchanges to gain unauthorized access.
    \item \textbf{Replay} other clients’ message exchanges to gain unauthorized access.
\end{compactitem}
We do not consider denial-of-service (DoS) attacks whereby \sadv floods the device with fake requests (tickets).
Techniques such as \cite{muraleedharan2006jamming, zhijun2020low, mamdouh2018securing} can mitigate these attacks.
We also do not consider physical attacks whereby \sadv physically tampers with devices via inducing hardware 
faults or modifying code in RoT. We refer to \cite{ravi2004tamper,obermaier2018past} for an overview of countermeasures 
against such attacks. Finally, we also do not consider side-channel attacks, similar to the Kerberos threat model.

\subsection{Security Requirements}
\label{sec: security requirements}
In order to be robust against \sadv defined in Section \ref{sec: adversary model}, \system needs to satisfy the following requirements:
\begin{itemize}
    \item Realize secure and efficient authenticated access control to resource-constrained IoT devices.
    \item Confirm the integrity of the device software state.
    \item Prevent impersonation of an authenticated client.
    \item Guarantee confidentiality of sensitive information such as session keys during transmission over the network.
    \item Ensure unforgeability and freshness of messages exchanged between two mutually authenticated entities.
\end{itemize}

%% file: content/04-protocols.tex
\section{\system Protocols}
\label{sec: main protocol}

Recall that \iotserver is treated as any other application server. 
The steps to obtain a ticket for \iotserver are the same as obtaining a Kerberos ticket. Hence, we focus on the steps starting from \iotserver issuing a ticket to a client for 
IoT device access.

\iotserver maintains a database with each device's ID, device type, long-term secret key(s), 
access control policy, and current synchronization value (in case of \pcd).

Protocol notation is summarized in Table \ref{table:notations}. 
One term that needs further clarification is \csync: it is 
the counter synchronized between \iotserver and the device, which keeps 
track of the number of synchronization requests sent by the device so far. 
It is used by \iotserver to authenticate the device during the synchronization phase. 
Therefore, the device needs to store \csync in non-volatile (persistent) memory. 
To simplify protocol description, we use 'K' to denote all long-term secret keys. 
However, in reality, three distinct long-term keys are shared between \iotserver 
and each device. One key is used for ticket generation and verification, 
another -- for session key generation, and the third -- for synchronization.

\subsection{General Device Protocol}\label{sec: general devices}
\begin{figure}
    \centering
    \includegraphics[width=0.5\textwidth]{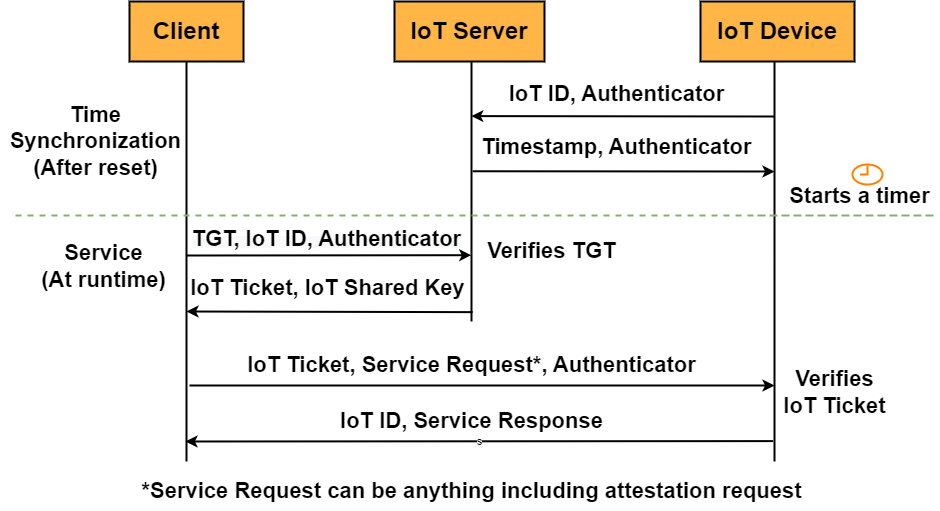}
    \caption{\boldmath{\gd} Protocol Overview.}     
    \label{fig:always-on overview}
\end{figure}

Figure \ref{fig:always-on overview} provides an overview of \system protocol for 
\gd, while detailed description is presented in Figure \ref{fig:protocol-general}.

\begin{figure}
  \begin{mdframed}[linewidth=1pt]
    \begin{protocol}\label{protocol:protocol-normal}
      \footnotesize
        General Device Protocol\\
        \textbf{Time Synchronization Phase (\boldmath{\gd} $\leftrightarrow$ \iotserver)}
        \begin{compactenum}
            \item After booting up \gd increases persistent sync counter \csync value by 1 and computes authenticator \axy{\plaingd}{\iotserver}:
                \begin{equation} \label{eqn: IOT Auth}
                    HMAC (\pklxysync{\iotserver}{\plaingd},[\pidx{\plaingd} || \pcsync])
                \end{equation}
            \item Then \gd builds \reqxy{\plaingd}{\iotserver} and sends it to \iotserver:\\ \reqxy{\plaingd}{\iotserver} $=$ $\pidx{\plaingd} || \pcsync|| \paxy{\plaingd}{\iotserver}$
            \item \label{list: verify c sync} \iotserver verifies \csync. It also verifies \axy{\plaingd}{\iotserver} using Equation \ref{eqn: IOT Auth}. Upon successful verification, \iotserver updates local \csync with received \csync.
            \item Then \iotserver computes authenticator \axy{\iotserver}{\plaingd}:
                \begin{equation} \label{eqn: ISRV Auth}
                     HMAC(\pklxysync{\iotserver}{\plaingd},[\pidx{\iotserver} || \pcsync || \ptsxy{\iotserver}{\plaingd}])
                \end{equation}
            \item \label{list: timestamp in sync res} Finally \iotserver builds \resxy{\iotserver}{\plaingd} and sends it to \gd:\\ $\pidx{\iotserver} || \pcsync || \ptsxy{\iotserver}{\plaingd} || \paxy{\iotserver}{\plaingd}$
            \item \label{list: verify sync res auth} \gd verifies \axy{\iotserver}
            {\plaingd} using Equation \ref{eqn: ISRV Auth}. Then it updates start\_time with received \tsxy{\iotserver}{\plaingd} and starts a timer.
        \end{compactenum}
        \textbf{Ticket Issuing Phase (C $\leftrightarrow$ \iotserver)}
        \begin{compactenum}
            \item C performs the steps of Kerberos protocol to obtain Kerberos service ticket, \tx{\iotserver} and shared key \kxy{C}{\iotserver} for \iotserver.  
            \item Computes $\paxy{C}{\iotserver} = E (\pkxy{C}{\iotserver}, [\pidx{C} || \padc || \ptsxy{C}{\iotserver}])$
            \item Builds ticket request \reqxy{C}{\iotserver} and sends it to \iotserver:\\ \reqxy{C}{\iotserver} $=$ $\pidx{\plaingd} || \ptx{\iotserver} || \paxy{C}{\iotserver}$
            \item \iotserver decrypts \tx{\iotserver} by using \klxy{\iotserver}{TGS} and verifies it. Then \iotserver issues IoT ticket \tx{\plaingd} and session key \kxy{C}{\plaingd} using different long term keys: 
                \begin{equation}
                \label{eqn: ticket session key GD}
                    HMAC (\pklxytkt{\plaingd}{\iotserver}, [\pidx{C} || \padc || \plfn{6} || \pidx{\plaingd}])
                \end{equation}
            \item Finally \iotserver builds response \resxy{\iotserver}{C} and sends it to C:
            \begin{equation*}
                E(\pkxy{C}{\iotserver},[\pidx{\plaingd} || \pkxy{C}{\plaingd} || \ptsxy{\iotserver}{C} || \plfn{6} || \ptx{\plaingd}])
            \end{equation*}

            \item \label{list: iot gd ticket} C decrypts \resxy{\iotserver}{C} with \kxy{C}{\iotserver}, retrieves and caches \kxy{C}{\plaingd}, \tx{\plaingd}.
          \end{compactenum}
        \textbf{Service Phase (C $\leftrightarrow$ \boldmath{\gd})}
        \begin{compactenum}
            \item \label{list: client gd auth} C computes authenticator \axy{C}{\plaingd}: 
                \begin{equation}\label{eqn: authenticator to GD}
                    \paxy{C}{\plaingd} = HMAC(\pkxy{C}{\plaingd}, [\ptsxy{C}{\plaingd}])
                \end{equation}
            \item Then C builds \reqxy{C}{\plaingd} and sends it to \gd:
            \begin{equation*}
            serv\_req || \pidx{C} || \padc || \plfn{6} || \ptx{\plaingd} || \ptsxy{C}{\plaingd} || \paxy{C}{\plaingd}
            \end{equation*}
            \item \label{list: plaintext timestamp} \gd checks plaintext \tsxy{C}{\plaingd} and \lfn{6} values, generates \kxy{C}{\plaingd} locally and verifies \tx{\plaingd} using Equation \ref{eqn: ticket session key GD}, verifies \axy{C}{\plaingd} using Equation \ref{eqn: authenticator to GD}.
            \item Finally \gd performs the requested service and sends back either plaintext service\_response (non-sensitive information) or service\_response encrypted with \kxy{C}{\plaingd} (sensitive information).
            \item C receives service\_response and decrypts with \kxy{C}{\plaingd} if necessary.
          \end{compactenum}
      \end{protocol}
  \end{mdframed}
  \caption{\boldmath{\gd} Protocol.}
  \label{fig:protocol-general}
\end{figure}

\textbf{Time Synchronization Phase:}
Time Synchronization Phase of Protocol \ref{protocol:protocol-normal} approximates 
wall-clock time by using a timer. After reboot, \gd obtains the current timestamp 
from \iotserver, stores it as its start\_time, and starts a timer.

Upon receiving a synchronization request \reqxy{\plaingd}{\iotserver}, \iotserver verifies it. 
In step \ref{list: verify c sync}, \iotserver verifies \csync by checking that it either equals local \csync for \gd or exceeds it by 1. 
\csync is included in the authenticator \axy{\plaingd}{\iotserver} in order to prevent 
replayed synchronization requests. Ideally, received \csync should be greater than the 
local version by 1. However, it might be equal to the local version 
due to lost response and subsequent re-transmission. \iotserver also verifies authenticator 
\axy{\plaingd}{\iotserver}. If both \csync and \axy{\plaingd}{\iotserver} are valid, 
then \gd is considered authenticated and \iotserver replies with a 
synchronization response -- \resxy{\iotserver}{\plaingd}. 
 
After receiving \resxy{\iotserver}{\plaingd}, \gd authenticates 
\iotserver in step \ref{list: verify sync res auth} by verifying \axy{\iotserver}{\plaingd}. Since \csync grows monotonically, it also acts as a nonce in 
computing \axy{\iotserver}{\plaingd}. This prevents replays of old synchronization responses. 
If verification succeeds, \gd stores \tsxy{\iotserver}{\plaingd} 
as its start\_time. After synchronization of start\_time, 
local time is set to: start\_time plus current timer value. 
If \gd's timer drifts drastically, the calculated timestamp would not be 
reliable. In such cases, it is recommended that this protocol is executed 
not only at boot time but also at regular (long-term) intervals.

\textbf{Ticket Issuing Phase:}
A user wishing to request a service from \gd must obtain an IoT ticket from \iotserver 
following the steps described in the Service phase of Protocol \ref{protocol:protocol-normal}. 
This protocol includes ticket lifetime \lfn{6} in IoT tickets to maintain ticket validity 
periods. This enables \iotserver to grant tickets to multiple clients for the same time period. 
Also, each ticket can be used multiple times before its expiration time. 
IoT ticket for \gd is generated by computing an HMAC over (\idx{C}, \adc, \lfn{6}, 
\idx{\plaingd}) with a long-term key. 
The corresponding session key is generated by computing HMAC over the same values 
with a {\it different} long-term key. 
The IoT ticket generation process is different from that in Kerberos. 
The latter is secured using encryption, while IoT tickets are secured via HMAC.

\textbf{Service Phase:}
This part of the protocol is used by users to request a service from a \gd. Since IoT tickets 
are multi-use, replay attacks pose a problem. To mitigate them, in step 
\ref{list: client gd auth} of the Service Phase, C calculates an authenticator 
\axy{C}{\plaingd} by computing HMAC over its current timestamp \tsxy{C}{\plaingd}
with a session key \kxy{C}{\plaingd}. Then, C includes both \tsxy{C}{\plaingd} and 
\axy{C}{\plaingd} in its service request, \reqxy{C}{\plaingd}. Together 
with \tsxy{C}{\plaingd}, \axy{C}{\plaingd} mitigates replays.

Upon receiving \reqxy{C}{\plaingd}, \gd computes the local timestamp by adding the 
current timer value to start\_time. Then, in step \ref{list: plaintext timestamp}, 
\gd checks if the difference between plaintext \tsxy{C}{\plaingd} and the
local timestamp is within a predefined short range. If not, it discards the 
request and sends back an “Invalid Request” response. \gd{} also checks if plaintext ticket lifetime \lfn{6} is later than the current local time. If not, it discards the request 
and sends back a “Ticket Expired” response. Next, \gd verifies \axy{C}{\plaingd}. Finally, \gd verifies the integrity of the IoT ticket 
by computing an HMAC over plaintext (\idx{C}, \adc, \lfn{6}, \idx{\plaingd}) 
with its long-term key and comparing it with the corresponding
value in the received ticket. This integrity check prevents the adversary from modifying 
the expiration time with the purpose of extending the ticket lifetime.

\noindent \textbf{Attestation as a Service:} Along with standard IoT device functionalities, 
\gd makes its own attestation available as a service. Any user with a valid 
IoT ticket can act as a verifier and request \gd to attest itself. Upon receiving an 
attestation request and validating the IoT ticket, RoT inside \gd calculates an 
HMAC over the program memory using the session key \kxy{C}{\plaingd}. 
\gd sends back the computed attestation result (HMAC) to the user. 
Acting as a verifier, the user knows the expected HMAC value for the benign 
(expected) software state of \gd  and thus can verify the response to determine 
\gd's current software state.

\subsection{Protocol for Power Constrained Devices}
\label{sec: power constraint devices}
\begin{figure}
    \centering
    \includegraphics[width=0.5\textwidth]{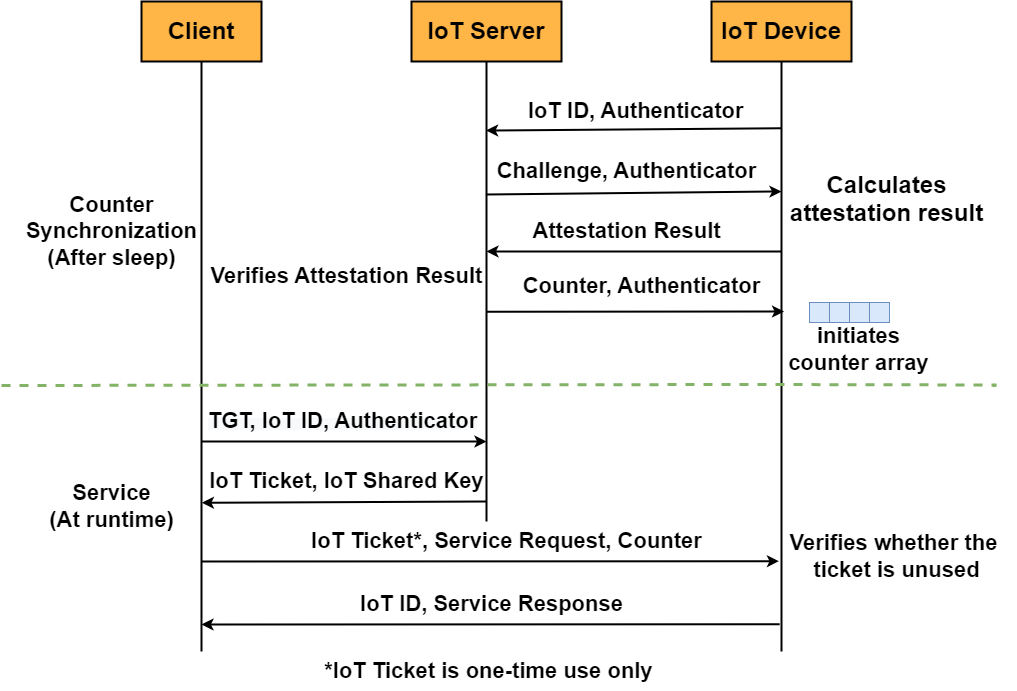}
    \caption{\boldmath{\pcd} Protocol Overview.} 
    \label{fig:nordic overview}
\end{figure}

Figure \ref{fig:nordic overview} provides an overview of \system protocol for \pcd, 
while detailed description is presented in Figure \ref{fig:protocol-power-constrained}.

\begin{figure}
    \captionsetup{justification=centering}
    \begin{mdframed}[linewidth=1pt]
    \begin{protocol}\label{protocol:protocol-nordic}
    \footnotesize
        Power constrained Device Protocol\\
        \textbf{Counter Synchronization Phase (\boldmath{\pcd} $\leftrightarrow$ \iotserver)}
        \begin{compactenum}
            \item Upon \pcd waking up, \pcd and \iotserver perform the same steps as steps 1 to 3 from the Time Synchronization Phase of Protocol \ref{protocol:protocol-normal} to build and verify \reqxy{\plainpc}{\iotserver} respectively. 
            \item Then \iotserver generates a random number as challenge and computes authenticator \axyattest{\iotserver}{\plainpc}:
            \begin{equation}\label{eqn: ISRV Auth Attest}
            HMAC(\pklxysync{\plainpc}{\iotserver},[\pidx{\iotserver} || challenge])
            \end{equation}
            \item \iotserver builds attestation request, \reqattest and sends it to \pcd: 
             $\pidx{\iotserver} || challenge || \paxyattest{\iotserver}{\plainpc}$
            \item \label{list: attestation key} \pcd verifies \axyattest{\iotserver}{\plainpc} using Equation \ref{eqn: ISRV Auth Attest} and generates Attestation Key, \kxyattest: 
            \begin{equation}\label{eqn: attestation key}
            \pkxyattest = HMAC(\pklxykey{\plainpc}{\iotserver},[challenge])
            \end{equation}
            \item \pcd calculates HMAC of memory region and sends it to \iotserver: 
            \begin{equation}\label{attestation hmac}
            Attst_{hmac} = HMAC(\pkxyattest, [Memory])
            \end{equation}
            \item \label{list: initialize counter} Upon receiving $Attst_{hmac}$, \iotserver generates \kxyattest using Equation \ref{eqn: attestation key} and verifies $Attst_{hmac}$ using Equation \ref{attestation hmac}. It also assigns current timestamp value \tsxy{\iotserver}{\plainpc} to \cpcd.
            \item Then \iotserver and \pcd perform the same steps as steps 4 to 6 from the Time Synchronization Phase of Protocol \ref{protocol:protocol-normal} to build and verify \resxy{\iotserver}{\plainpc} respectively. 
            \item \pcd initializes local counter array using \tsxy{\iotserver}{\plainpc}.
        \end{compactenum}

    \textbf{Ticket Issuing Phase (C$\leftrightarrow$\iotserver)}
        \begin{compactenum}
            \item C performs the steps of Kerberos protocol to obtain Kerberos service ticket \tx{\iotserver} and shared key \kxy{C}{\iotserver} for \iotserver. 
            Then C and \iotserver perform the same steps as steps 2 to 4 from the Ticket Issuing phase of Protocol \ref{protocol:protocol-normal} to build and verify \reqxy{C}{\iotserver} respectively.
            \item Then \iotserver issues IoT ticket \tx{\plainpc} and generates session key \kxy{C}{\plainpc} using different long term keys: 
            \begin{equation}\label{eqn: ticket PC}
                HMAC (\pklxytkt{\plainpc}{\iotserver}, [\pidx{C} || \padc || \pcpcd || \pidx{\plainpc}])
            \end{equation}
            \item Finally \iotserver builds \resxy{\iotserver}{C} and sends it to C:
             \begin{multline*}
                 E(\pkxy{C}{\iotserver},[\pidx{\iotserver} || \pkxy{C}{\plainpc} || \ptsxy{\iotserver}{C} || \pcpcd ||\\ \ptx{\plainpc}])
             \end{multline*}
            
            \item \label{list: iot pc ticket} C decrypts \resxy{\iotserver}{C} with \kxy{C}{\iotserver}, retrieves and caches \kxy{C}{\plainpc}, \tx{\plainpc}.
        \end{compactenum}
    \textbf{Service Phase (C$\leftrightarrow$\boldmath{\pcd})}
        \begin{compactenum}
            \item C builds \reqxy{C}{\plainpc} and sends it to \pcd:\\
            $service\_request || \pidx{C} || \padc || \pcpcd || \ptx{\plainpc}$
            \item \label{list: plaintext counter} \pcd checks plaintext \cpcd value, generates \kxy{C}{\plainpc} locally, and verifies \tx{\plainpc} using Equation \ref{eqn: ticket PC}.     
            \item Finally \pcd performs the requested service and sends back either plaintext service\_response (non-sensitive information) or service\_response encrypted with \kxy{C}{\plainpc} (sensitive information).
            \item C receives service\_response and decrypts with \kxy{C}{\plainpc} if necessary.
        \end{compactenum}
    \end{protocol}
  \end{mdframed}
  \caption{\boldmath{\pcd} Protocol. 
  }
  \label{fig:protocol-power-constrained}
\end{figure}

\textbf{Counter Synchronization Phase:}
\iotserver maintains a separate synchronization value for each \pcd, and this value is 
reinitialized with the current timestamp from the local clock of \iotserver 
every time \pcd performs synchronization.

When \pcd{} boots up, it follows the steps outlined in Counter Synchronization Phase 
of Protocol \ref{protocol:protocol-nordic} to obtain the synchronization value from \iotserver. 
\iotserver also requests an attestation report from \pcd{} during this process.
Only if that report is valid, i.e., if \pcd is healthy, \iotserver grants tickets to 
clients to access \pcd. 
This eliminates the need to expose attestation as a service for \pcd, 
given that \pcd is low in runtime and power budget.

\textbf{Ticket Issuing Phase:}
This part of Protocol \ref{protocol:protocol-nordic} describes how clients obtain IoT Tickets 
from \iotserver, prior to to requesting service from \pcd. These tickets are single-use only. 
The rationale is that synchronizing and running a timer in \pcd{} for a short period 
of activity is expensive and offers low utility. 
As a result, tickets can not include a timestamp to indicate their validity period. 
Instead, \iotserver maintains a separate counter initialized with the synchronization 
value for each \pcd. Recall that the synchronization value is the timestamp sent to 
\pcd during Counter Synchronization Phase. Each time \iotserver receives a ticket 
request from a client, it increases the counter value by 1 and uses the counter value 
as the nonce in the ticket, ensuring the ticket is fresh and unique for a single use.

\textbf{Service Phase:}
In this phase, clients use Protocol \ref{protocol:protocol-nordic} to request service 
from \pcd. However, nonce-based \pcd{} tickets are prone to race conditions. For example,
suppose that Client-A and Client-B both obtain IoT tickets from \iotserver. 
However, Client-A obtains its ticket before Client-B, resulting in the former's 
ticket having a lower-numbered nonce than the ticket of the latter.  
(Recall that we treat nonces as monotonic counters.)
If Client-B presents its ticket to the device first, \pcd updates its local
counter value with the nonce of Client-B's ticket. In that case, Client-A becomes
unable to use its ticket since the nonce in Client-A's ticket is lower than the current 
counter value of \pcd.

To avoid such anomalies, \pcd{} maintains a {\it counter buffer} of size \textit{n}, where
{\it n} is the maximum number of clients that can obtain tickets for \pcd{} during each 
of its liveness (awake) period.  
In step \ref{list: plaintext counter} of the Service Phase of 
Protocol \ref{protocol:protocol-nordic}, when \pcd{} receives a ticket, it checks whether its 
nonce is within the counter buffer and is still unused. If this check fails, \pcd{} discards 
the request and returns an "Invalid Counter" response. 
This prevents race conditions for single-use \pcd tickets while keeping them non-blocking,
meaning that \iotserver does not reserve a specific time period for a client's \pcd ticket,
and does not reject tickets for other clients during that period.

%% file: content/05-implementation.tex
\section{Implementation Details} \label{sec:implementation}
This section describes the prototype implementation
of \iotserver, \gd, \pcd, and client application. All
source code is available at \cite{kesic-repo}.

\subsection{IoT Server}
We implemented \iotserver in Python. It has two main components: 

\textbf{Ticket Manager (\tktmanager)} is implemented as a web application using Flask library. 
It is hosted in Apache Web Server and configured for Kerberos Authentication.
A client needs to obtain a valid ticket from TGS to call \tktmanager.
\tktmanager follows a static policy to grant IoT tickets: 
it maintains a list of allowed users. A granted IoT ticket for a given
device is valid for all available functionalities of that device.

\textbf{Synchronization Manager (\syncmanager)} is implemented in Python. 
Two always-listening UDP server sockets are used for communication with devices. 
The first is responsible for accepting synchronization requests from devices, while
the second accepts attestation reports from \pcd as part of the synchronization process. 
Communication between \syncmanager and devices is protected by the long-term secret key, \klxysync{\plaingd}{\iotserver} or \klxysync{\plainpc}{\iotserver}, 
shared between \iotserver and each device.

\syncmanager runs as a separate thread from \tktmanager.
\iotserver is hosted on a Linux laptop with Intel(R) Core(TM) i7-8550U 
CPU running at 1.80GHz, with 8GB RAM.

\subsection{\gd and \pcd}
An NXP LPCXpresso55S69 development board with TrustZone-M emulates an IoT device. 
The board is based on ARM Cortex-M33 MCU. It runs at 150 MHz with 640KB flash and 320KB SRAM. 
Wifi 10 click board is used -- along with LPCXpresso55S69 board -- for WiFi connectivity. 
\gd and \pcd are emulated separately. The former is a smart bulb where clients 
control the following features: (1) turning on LED, (2) turning off LED, and
(3) performing attestation. The latter is also a smart bulb, however, clients 
can only turn on and turn off LED. It does not support attestation. 

The program running on the emulated device is divided into two parts:

\textbf{Non-Secure Part:} processes user commands. It is also responsible for 
the synchronization process with \iotserver. Network communication and 
actuation (turning on/off LED) are handled by this part. All communication 
is over UDP.
 
\textbf{Secure Part:} works as an RoT. It stores secret keys and synchronization value, 
as well as timer/counter value. It is responsible for all cryptographic operations, 
i.e., HMACs, and uses the Mbed-TLS library.

These two parts are compiled into separate .axf (binary) files and also 
flashed to the board separately. Moreover, they are executed using separate RAM.

\subsection{Client Application}
A sample client application is written in Python. It obtains IoT tickets from 
\iotserver and requests service from \gd and \pcd. Before the client application 
calls \tktmanager to request an IoT ticket, the client (user) must obtain a Kerberos 
ticket for \iotserver. This is done by configuring a Kerberos client on the client
machine and calling kinit. Then, the client application automatically 
includes the Kerberos ticket in the request header when it calls \tktmanager. 

\subsection{Remote Attestation Process}
Both \gd and \pcd can perform remote attestation. 
\pcd performs attestation during the synchronization phase with \iotserver 
acting as the verifier. In case of \gd, a user (acting as a verifier) can 
ask \gd to perform attestation during the service phase. 
In both cases, the attestation process is the same. 

The RoT in the secure part of the device computes a hash of the entire non-secure 
program flash memory using \textit{mbedtls\_sha256}. Then, the appropriate key is 
used to compute an HMAC over this hash value using \textit{MBEDTLS\_MD\_SHA256}. 
Note that, for \gd, this attestation key is the service session key, \kxy{C}{\plaingd}, 
shared with client (C). For \pcd, it is a temporary attestation key, 
\kxyattest computed during the synchronization process. 

The computed HMAC  is sent to the verifier. 
The correct (expected) reference hash for a given device is assumed to correspond to
the latest legitimate software version that the verifier expects the device to run. 
Since the verifier is assumed to know both the attestation key and the expected reference 
hash, it computes its own expected HMAC and compares it with that 
received from the device, which allows to determine whether the device is malware-free.

\subsection{Protocol Message Details} \label{appendix: protocol messages}
This section details the format and length of various request and response messages exchanged between client-\iotserver, \iotserver-device and client-device. 

\begin{figure}
    \centering
    \includegraphics[width=\linewidth]{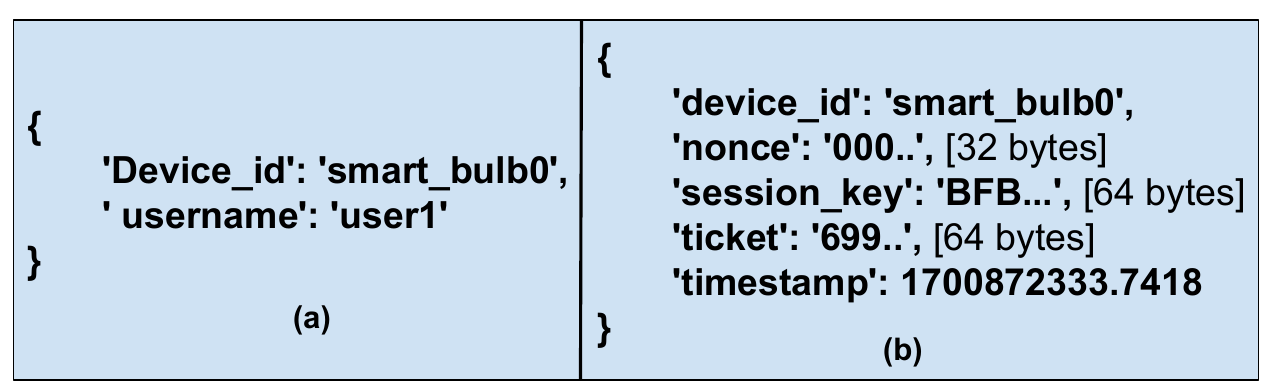}
    \caption{(a) IoT Ticket Request from Client to \iotserver, (b) IoT Ticket Response from \iotserver.}   
    \label{fig: ticket req res}
    \vspace{-.4cm}
\end{figure}

\noindent \textbf{Between Client \& \iotserver}
Clients send requests in ``application/json'' format to \iotserver 
and set two properties: user\_name, device\_id. 
There is no length constraint for these fields. 
The responses from \iotserver are also in ``application/json'' format, and they include five properties: device\_id, nonce, session\_key, ticket, and timestamp. The actual type of nonce depends on device type. For \gd devices, the nonce is the lifetime of the ticket. For \pcd devices, it is the counter value. Figure \ref{fig: ticket req res} provides an example of the request and response. 
The user name and device ID provided by the client here are not the numerical client ID and device ID used in generating the ticket. \iotserver maintains a mapping from semantic user name to numerical client ID and a similar mapping from semantic device ID to numerical device ID.

\begin{figure}
    \centering
    \includegraphics[width=\linewidth]{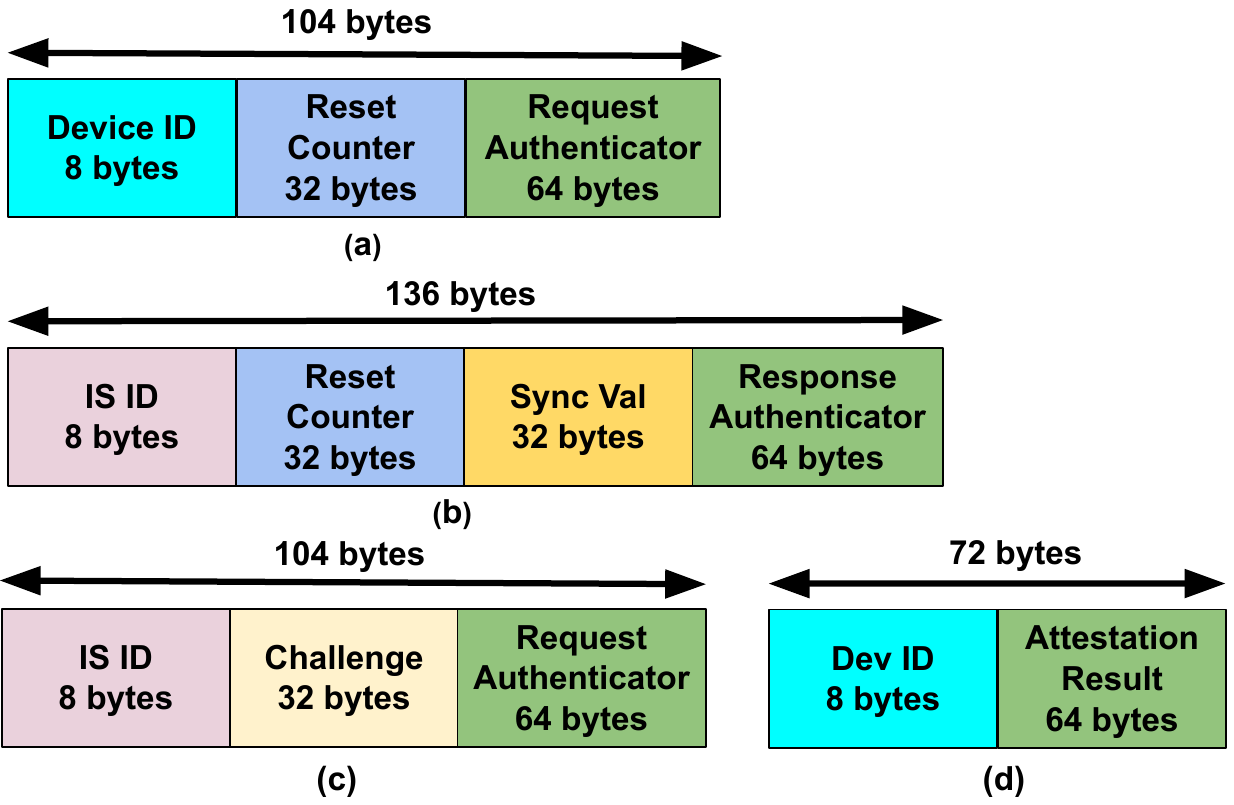}
    \caption{(a) Synchronization Request from IoT Device to \iotserver, (b) Synchronization Response from \iotserver to IoT device, (c) Attestation Request from \iotserver to \boldmath{\pcd}, (d) Attestation Response from \boldmath{\pcd} to \iotserver.}   
    \vspace{-.2cm}
    \label{fig: sync attest req res}
    \vspace{-.4cm}
\end{figure}

\noindent \textbf{Between \iotserver \& Device}
All request and response messages exchanged between \iotserver and IoT devices are in string format. 
The whole string is made up by concatenating multiple fixed-length fields in a specific serial. 
The synchronization request and response format are the same for both \gd and \pcd. ``Sync Val'' field in synchronization response contains the current timestamp value from \iotserver. Additionally, \pcd exchanges attestation request and response with \iotserver. Figure \ref{fig: sync attest req res} provides the request and response formats.

\begin{figure}
    \centering
    \includegraphics[width=\linewidth]{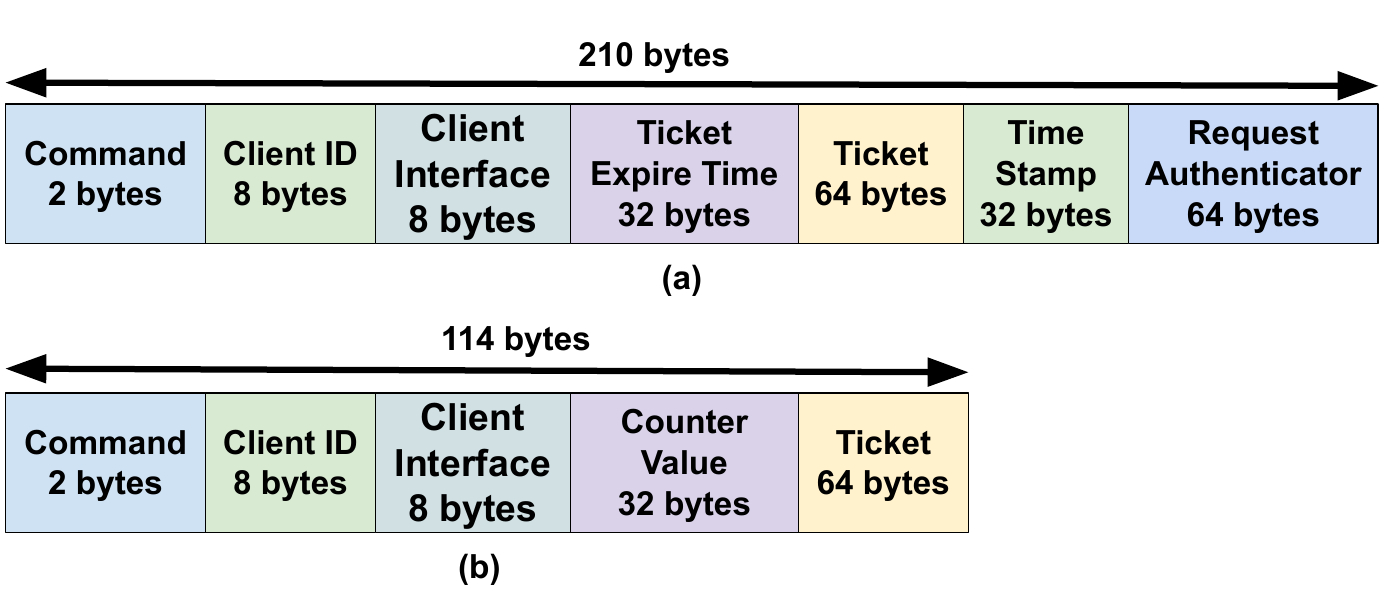}
    \caption{(a) Service Request from Client to \boldmath{\gd}, (b) Service Request from Client to \boldmath{\pcd}.}     
    \label{fig: gd pcd service req}
    \vspace{-.4cm}
\end{figure}

\noindent \textbf{Between Client \& Device}
The service requests from the client to \gd and \pcd follow different formats. This is due to the differing ticket format and capabilities of \gd and \pcd. We provide these formats in Figure \ref{fig: gd pcd service req}. There is no fixed format for service response since it depends on the command.

%% file: content/06-evaluation.tex
\section{Evaluation}
This section presents security and performance analyses of \system. 
As part of performance analysis, we evaluate \system in terms of storage, 
memory, run time, and network overhead.  Since \iotserver and client 
applications are expected to run on powerful devices, overheads of these 
components are minimal and not discussed here due to space constraints. 
Thus, we focus on the overhead of IoT devices. 

\subsection{Security Analysis}
We now argue that \system meets security requirements in 
Section \ref{sec: security requirements}.

Recall that \iotserver authenticates users (clients), manages access policy, 
and grants tickets to access devices. A client does not authenticate to 
each IoT device and an IoT device does not authenticate each client. 
\system uses a timer-based ticket format for \gd{}, while a 
counter-based simple ticket format is used for \pcd{}.

\system supports attestation on both \gd and \pcd as long as an RoT is present. 

At login time, Kerberos AS authenticates each client (C). We thus assume 
that client impersonation is not possible.

Communication between C and \iotserver is encrypted with the session key 
shared among them. As a result, no eavesdropper can extract the session 
key(\kxy{\plaingd}{\iotserver}, or \kxy{\plainpc}{\iotserver}) 
in the ticket response from \iotserver to C. This session key 
is intended to secure the communication between C and the device. 

Requests from C to \iotserver are encrypted. Tampering with encrypted requests 
results in failed decryption and is detectable. On the other hand, plaintext 
requests from clients to devices include an HMAC of the original request, which 
is computed using a secret session key known only to C and the device. The latter
detects tampered requests by recomputing the HMAC.

For multi-use tickets (\tx{\plaingd}), C provides an authenticator
along with each request. An authenticator is computed using a secret 
session key shared between C and the device. Computation of an 
authenticator includes the current timestamp, which precludes replays of
authenticators.   

The above shows that \system meets the security requirements stated in 
Section \ref{sec: security requirements}.

\subsection{Storage \& Memory Overhead}
\begin{figure}
    \centering
    \includegraphics[width=0.5\textwidth]{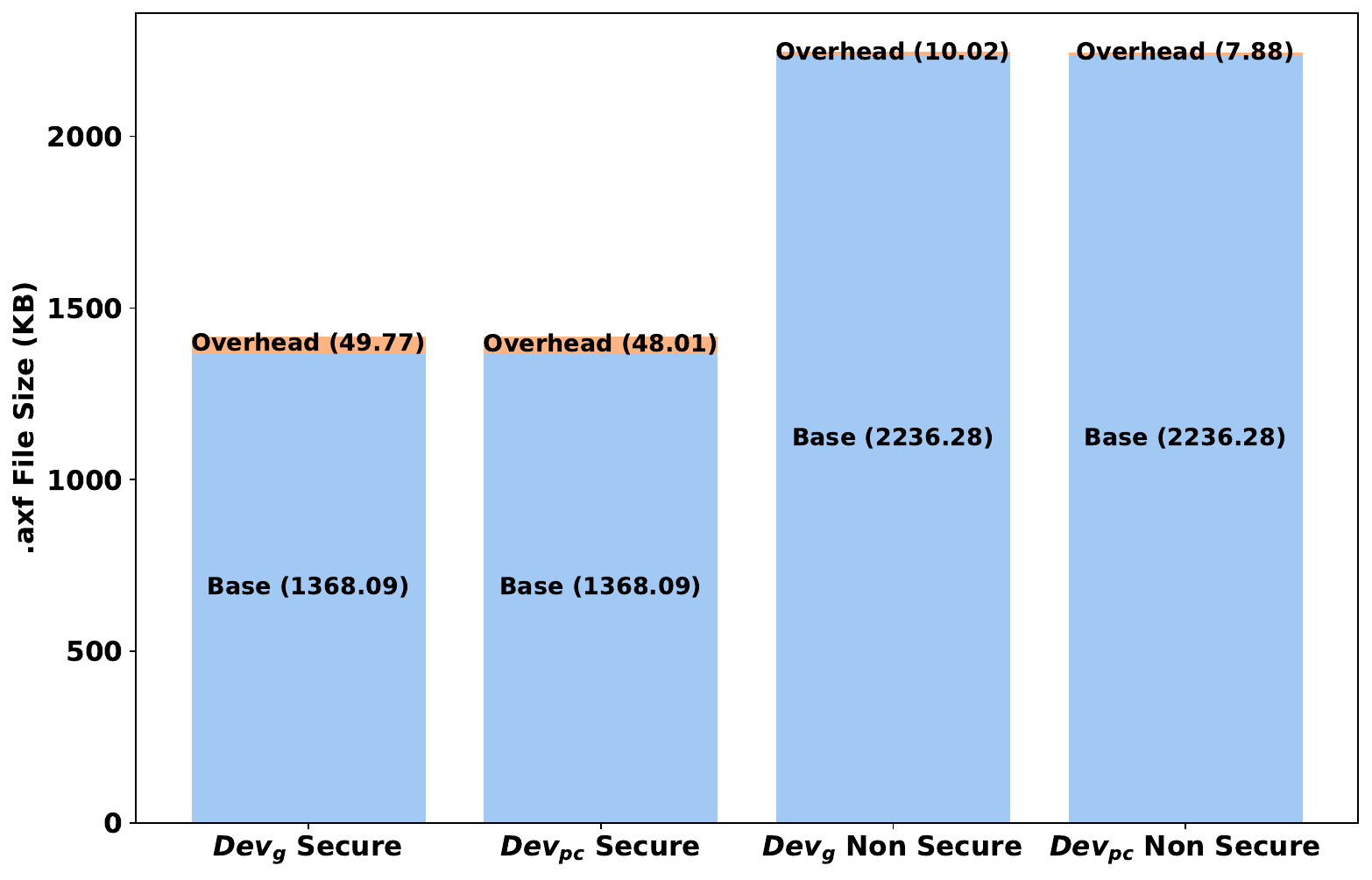}
    \caption{Storage Overhead for \boldmath{\gd} \& \boldmath{\pcd}.}
    \label{fig: storage overhead}
\end{figure}

\begin{figure}
    \centering
    \includegraphics[width=0.5\textwidth]{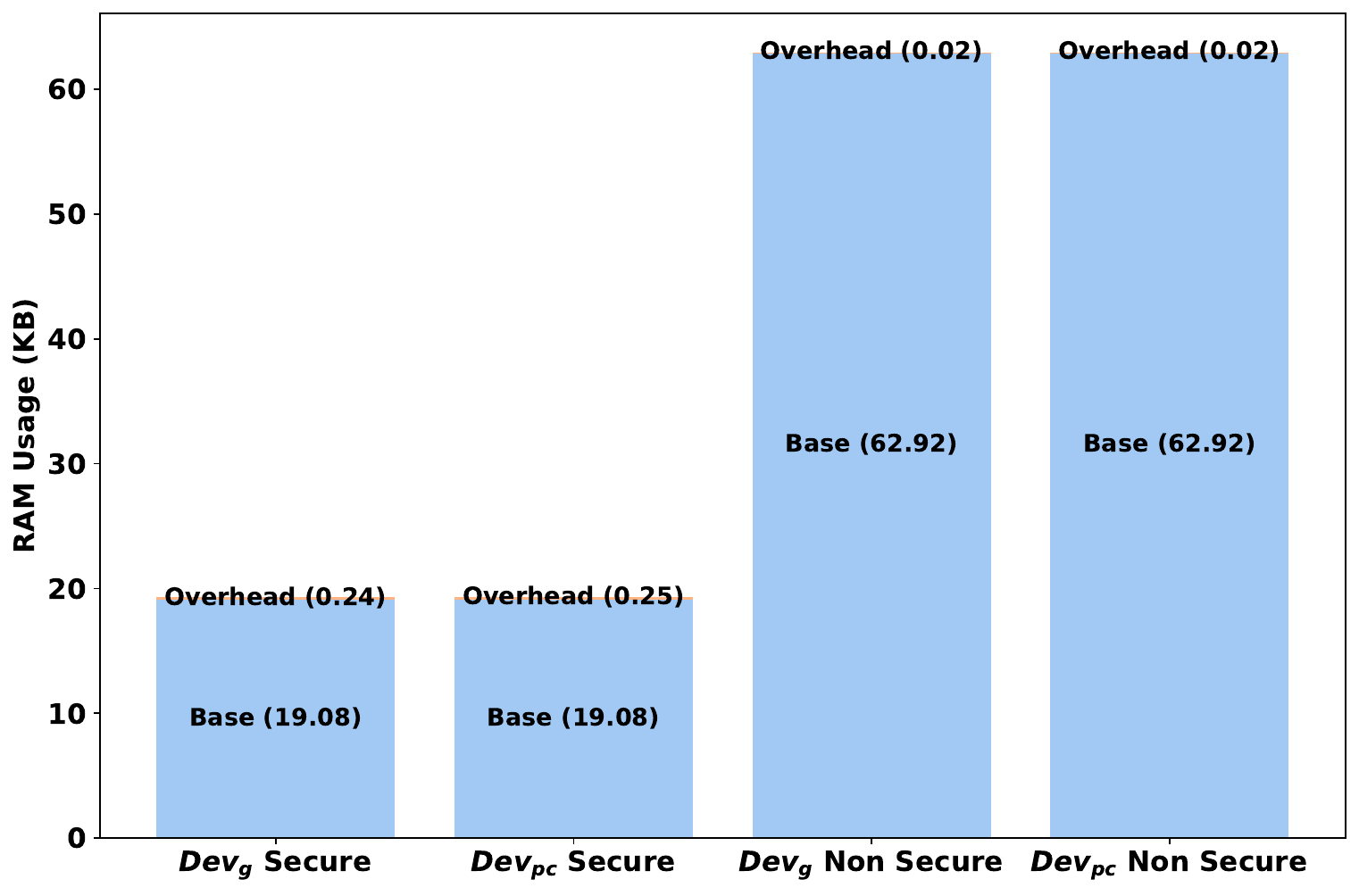}
    \caption{Memory Overhead for \boldmath{\gd} \& \boldmath{\pcd}.}
    \label{fig: memory overhead}
\end{figure}

We assess storage and memory \system overhead for 
\gd and \pcd prototype implementations. 
We calculate the overheads separately for secure part and non-secure part. Figure \ref{fig: storage overhead} and Figure \ref{fig: memory overhead} show the details about storage and memory overhead, respectively.

We measure storage overhead in terms of the .axf file size. 

Memory overhead is measured by the increase in RAM usage. 
As evident from Figure \ref{fig: storage overhead} and Figure \ref{fig: memory overhead},
storage and memory overheads 
are very low for both device types. \system causes only a 60KB increase overall ((considering both secure and non-secure parts)
for \gd over .axf file sizes of 3604 KB. For \pcd, this increase is even 
lower: 56KB. Storage overhead is 1.66\% for \gd and 1.55\% for \pcd. 
Furthermore, overall memory overhead is $\sim$0.3\% for both \gd and \pcd. 
This low overhead is due to the simple and lightweight implementation in \system. 

\begin{table}
    \caption{Storage \& Memory Overhead for \boldmath{\gd} \&  \boldmath{\pcd}.} 
  \footnotesize
  \centering
     \begin{tabular}{|c|r|r|r|} \hline
        Type & Base Device & \gd Overhead & \pcd Overhead \\ \hline
        Storage (KB)  &  3604 & 60 & 56 \\ \hline
        Memory (KB) & 82 & 256 & 272 \\ \hline
     \end{tabular}
  \label{tab: storage memory overheads}
\end{table}

\subsection{Runtime Overhead}
\gd and \pcd incur different runtime overheads in synchronization and service phases. 
We measure overhead in terms of cycles to complete individual steps within 
each phase and also the whole process. Using the operational frequency of the board, 
we also compute elapsed time. Note that we measure the mean and standard 
deviation of each performance value over 10 iterations.

\textbf{\boldmath{\gd} Synchronization Overhead:}
This phase involves three steps: preparing the request, sending it to \iotserver, and 
processing the response. The overhead for all steps (and cumulative) is 
presented in Table \ref{tab: timer sync overhead}. Steps 1 and 3 happen 
on the device; thus their standard deviations are negligible. However, step 2 includes 
processing time on \iotserver and network latency, which results in a higher standard deviation. 
The synchronization time is minimal at 44.1648 ms on average for \gd. Recall that it
only occurs at boot time.

\begin{table}
  \caption{Synchronization Runtime Overhead for \boldmath{\gd}.
  } 
  \centering
  \resizebox{\columnwidth}{!} {
     \begin{tabular}{|c|r|r|r|r|} \hline
     \multirow{2}{*}{Step}  & \multicolumn{2}{c|} {Kilocycles}  & \multicolumn{2}{c|} {Time @ 150MHz (ms)} 
     \\ \cline{2-5} &  
     Mean & Standard Deviation  & Mean & Standard Deviation
    \\ \Xhline{1\arrayrulewidth}
    Prepare Request & 288.6801 & 0.1459 & 1.9245 & 0.0009 \\ \hline
    Latency to Receive Response & 5684.7393 & 1047.2006 & 37.8982 & 6.9813 \\ \hline
    Process Response & 651.2986 & 0.3453 & 4.3419 & 0.0023\\ \hline
    \textbf{End-to-End Process} & 6624.7180 & 1047.1781 & 44.1648 & 6.9812\\ \hline
     \end{tabular}
  }
  \label{tab: timer sync overhead}
\end{table}

\textbf{\boldmath{\gd} Service Overhead:}
During service phase, \system verifies IoT tickets, which incurs runtime overhead that varies 
based on the service request scenario. Values of all ticket are checked one by one, and the ticket is 
rejected if any verification fails. Due to that, runtime overhead is highest when plaintext 
expiration time is modified, and lowest -- when plaintext timestamp value is wrong. The 
verification process takes 8.4487 ms (on average) for a valid ticket, as shown in Table 
\ref{tab: GD service overhead}. However, \system does not handle side-channel attacks 
that might occur due to timing differences; constant-time 
techniques \cite{constanttime96} should be used for mitigation.

\begin{table}
  \caption{Service Runtime Overhead on \boldmath{\gd}.} 
  \footnotesize
  \centering
  \resizebox{\columnwidth}{!} {
     \begin{tabular}{|c|r|r|r|r|} \hline
     \multirow{2}{*}{Scenario}  & \multicolumn{2}{c|} {Kilocycles}  & \multicolumn{2}{c|} {Time @ 150MHz (ms)} 
     \\ \cline{2-5} &  
     Mean & Standard Deviation  & Mean & Standard Deviation
    \\ \Xhline{1\arrayrulewidth}
    Wrong Time stamp & 35.8300 & 0 & 0.2389 & 0 \\ \hline
    Tampered Time stamp & 734.3302 & 0.4645 & 4.8955 & 0.0031 \\ \hline
    Expired Ticket & 990.0922 & 0.2942 & 6.6006 & 0.0019\\ \hline
    Valid/Tampered Ticket & 1267.3084 & 39.4993 & 8.4487 & 0.2633\\ \hline
     \end{tabular}
  }
  \label{tab: GD service overhead}
\end{table}
 
\textbf{\boldmath{\pcd} Synchronization Overhead:}
The synchronization phase of \pcd involves multiple communication rounds
and remote attestation. Thus, it incurs higher runtime overhead than \gd. However, 
it ensures that the device is healthy and frees the users from having to perform  
remote attestations separately.

\begin{figure}
    \centering
    \includegraphics[width=0.85\linewidth]{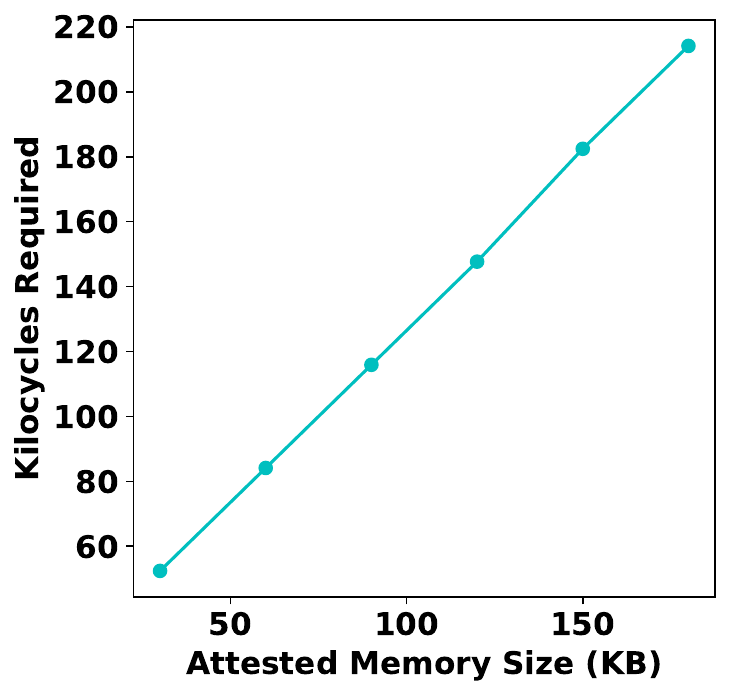}
    \caption{Attestation Overhead Based on Memory Size.}
    \label{fig: attestation overhead}
\end{figure}

There are five steps in this phase. Table \ref{tab: counter sync overhead} shows 
step-wise and cumulative runtime overhead. Steps 2 and 4 
include both processing time on \iotserver and network latency. They have 
higher standard deviations. Step 3 includes HMAC computation over non-secure flash memory region,
which depends on memory size and (unsurprisingly) overhead increases linearly with memory size. 
Figure \ref{fig: attestation overhead} presents the results on attestation overhead for varying memory sizes.

\begin{table}
  \caption{Synchronization Runtime Overhead for \boldmath{\pcd}.} 
  \footnotesize
  \centering
  \resizebox{\columnwidth}{!} {
     \begin{tabular}{|c|r|r|r|r|} \hline
     \multirow{2}{*}{Step}  & \multicolumn{2}{c|} {Kilocycles}  & \multicolumn{2}{c|} {Time @ 150MHz (ms)} 
     \\ \cline{2-5} &  
     Mean & Standard Deviation  & Mean & Standard Deviation
    \\ \Xhline{1\arrayrulewidth}
    Prepare Request & 295.7771 & 2.8463 & 1.9718 & 0.0189 \\ \hline
    Latency to Receive Attestation Request & 11745.8281 & 4537.9729 & 78.3055 & 30.2531 \\ \hline
    Process Attestation Request & 1010.9219 & 26.4953 & 6.7395 & 0.1766\\ \hline
    Latency to Receive Sync Response & 3092.8630 & 534.2839 & 20.6191 & 3.5619\\ \hline
    Process Sync Response & 677.6581 & 24.1696 & 4.5177 & 0.1611\\ \hline \hline
    End-to-End Process & 16823.04823 & 4695.3580 & 112.1537 & 31.3024\\ \hline
     \end{tabular}
  }
  \label{tab: counter sync overhead}
\end{table}

\textbf{\boldmath{\pcd} Service Overhead:}
Runtime overhead in the service phase is due to the ticket verification process, 
which is very fast for \pcd. Detecting a ticket with a wrong counter takes only 25.614 
kilocycles (0.1708 ms) on average, while validating a ticket takes 
541.532 kilocycles (3.6102 ms). Standard deviations are negligible in both scenarios. 
Details are shown in Table \ref{tab: PCD service overhead}. 

\begin{table}
\caption{Service Runtime Overhead on \boldmath{\pcd}.}
  \footnotesize
  \centering
  \resizebox{\columnwidth}{!} {
     \begin{tabular}{|c|r|r|r|r|} \hline
     \multirow{2}{*}{Scenario}  & \multicolumn{2}{c|} {Kilocycles}  & \multicolumn{2}{c|} {Time @ 150MHz (ms)} 
     \\ \cline{2-5} &  
     Mean & Standard Deviation  & Mean & Standard Deviation
    \\ \Xhline{1\arrayrulewidth}
    Wrong Counter Value & 25.6145 & 0.1990 & 0.1701 & 0.0013 \\ \hline
    Valid/Tampered Ticket & 541.5319 & 2.5956 & 3.6102 & 0.0173\\ \hline
     \end{tabular}
  }  
  \label{tab: PCD service overhead}
\end{table}

\subsection{Network Overhead}
We now consider network overhead on IoT devices caused by \system in synchronization and service phases. 
We consider sizes of all requests and responses exchanged by IoT devices. 
Network overhead is summarized in Table \ref{tab: network overheads}.

\begin{table}
    \caption{Network Overheads for \boldmath{\gd} and  \boldmath{\pcd}.} 
  \footnotesize
  \centering
  \scalebox{0.7}{\resizebox{\columnwidth}{!} {
     \begin{tabular}{|c|r|r|} \hline
     Phase & Device Type & Overhead (B) \\ \Xhline{1\arrayrulewidth}
     \multirow{2}{*}{Synchronization}  &  \gd & 240 \\ \cline{2-3} &  
     \pcd & 416\\ \Xhline{1\arrayrulewidth}
     \multirow{2}{*}{Service}  &  \gd & 208 \\ \cline{2-3} & \pcd & 112\\ \Xhline{1\arrayrulewidth}
     \end{tabular}
  }}
  \label{tab: network overheads}
\end{table}

Referring to Figure \ref{fig: sync attest req res}, synchronization request and response sizes are 104 and 136 bytes respectively. As a result, synchronization network overhead on \gd is 240 bytes in total. 
\pcd exchanges two additional messages with \iotserver: attestation request (104 bytes) and response 
(72 bytes), resulting in 416 bytes total. Similarly, network overhead during the service phase comes from service requests (Figure \ref{fig: gd pcd service req}). It is 208 bytes for \gd and 112 bytes for \pcd.

\subsection{Comparison with Kerberos} \label{subsec: comparison}
We compare the overhead of \system with that of Kerberos. However, as mentioned earlier in Section 
\ref{sec:design}, Kerberos can not be directly implemented on IoT devices. 
Therefore, we can not directly compare their respective overheads and instead
compare \system with standard Kerberos for its usual setting. Table \ref{tab:overheads comparison} shows 
the results. 

MIT Kerberos Repository \cite{mit_kerberos_repo} provides the implementation of the simplest Kerberos service application called ``sim\_server.c''. We use this application to calculate the minimum storage overhead of  Kerberos. The size of the compiled binary file is 35032 bytes. To be functional, this application requires a keytab file containing the service keys. Additionally, libkrb5-dev package must be installed in the system. The minimum size of a keytab file is 4000 bytes, whereas the size of libkrb5-dev installed on a Linux machine is 18200 bytes. Therefore, the storage overhead is (35032+182000+4000) bytes. 
This is a lower bound since ``sim\_server.c'' depends on other less important and small libraries as well.

Kerberos memory overhead can be calculated using MaxTokenSize parameter. It denotes the buffer size used to store authentication data, i.e., service ticket during runtime. The default value for MaxTokenSize was 12000 bytes in Windows 7 \cite{kerberos_buffer_size} and we consider that to be the minimal memory overhead of Kerberos. This is a conservative lower bound since that buffer size proved inefficient for some Kerberos tickets \cite{kerberos_buffer_size}. 
Consequently, the default value has been increased to 48000 bytes for Windows 8 and later machines.

Runtime overhead comes from the processing time (~7.8ms) of a Kerberos ticket for a 22000 MHz processor \cite{moralis2007performance}. 

Network overhead is calculated from the length of the authorization header that includes a Kerberos service ticket. We use our implemented client application for this purpose since it calls a REST API endpoint of \iotserver and authenticates itself by providing a Kerberos ticket in the authorization header.

As Table \ref{tab:overheads comparison} shows, storage, memory, runtime, and network overheads of \system are significantly 
lower than those of  Kerberos. \gd incurs 3.69$\times$ lower storage, 46.87$\times$ lower
memory, 135.41$\times$ lower runtime, and 4.07$\times$ lower, overhead. Meanwhile, 
\pcd exhibits 3.95$\times$ lower storage, 44.12$\times$ lower memory, 
316.88$\times$ lower runtime, and 7.55$\times$ lower network, overhead.

\begin{table}
    \caption{\boldmath{\system} vs Kerberos Overheads.} 
  \centering
  \resizebox{\columnwidth}{!} {
     \begin{tabular}{|c|r|r|r|} \hline
     Overhead & \gd & \pcd &  Kerberos \cite{mit_kerberos_repo, moralis2007performance, kerberos_buffer_size}\\ \Xhline{1\arrayrulewidth}
     Storage (KB) &  59.788 & 55.888 & 221.032 \\ \hline
     Memory (B) & 256 & 272 & 12000 \\ \hline 
     Runtime (ms) & 8.4 ms (@150 MHz)& 3.6 ms (@150 MHz) & 7.8ms (@2.2 GHz)\\ \hline
     Network (B) & 208 & 112 & 846\\ \hline
     \end{tabular}
  }
  \label{tab:overheads comparison}
\end{table}

%% file: content/07-related_works.tex
\section{Related Work}

\noindent \textbf{Kerberos-Based Authentication Schemes for IoT Devices:} 
Several efforts attempted to address the multi-user access problem for 
IoT devices by adopting Kerberos. There are two main directions in prior work: 

The first aims to decrease the computation and communication cost of Kerberos for IoT devices 
\cite{kerberos_iot,hardjono2014kerberos, aljanah2022multifactor,miyazawa2010design,tamboli2016secure, esfahani2017lightweight}. \cite{kerberos_iot} uses a nonce-based service ticket to grant access. 
However, the device can not verify the ticket locally and must communicate with KDC to do so.
\cite{aljanah2022multifactor} reduces the number of messages exchanged and the cost of constructing
a ticket. \cite{miyazawa2010design} uses table representation to reduce code size, memory copies, 
and heap allocations. 
None of these results, except \cite{kerberos_iot},
applies to IoT devices without real-time clocks.

The second direction focuses on addressing certain use-cases, such as machine-to-machine 
communication (communication among IoT devices) or using a central controller to manage all IoT devices \cite{gaikwad20153, 
esfahani2017lightweight, kim2017securing, thirumoorthy2022improved}. Such use-cases are 
different from what \system targets: allowing multiple users direct and secure access to  IoT devices. 
\cite{gaikwad20153} introduces a smart central controller to implement Kerberos for smart-home 
systems and maintains authentication and authorization at the controller level. 
\cite{esfahani2017lightweight} involves a low-cost Machine-to-Machine (M2M) protocol for
IoT devices to communicate with machines. \cite{thirumoorthy2022improved} uses 
an inter-server protocol: it establishes communication between two servers and 
uses an improved key agreement, as compared to  Kerberos. 

\textbf{Remote Attestation (RA):}
Prior RA schemes fall into three main categories based on the type of RoT: software-based, 
hardware-based, and hybrid. Software-based remote attestation \cite{seshadri2004swatt, pioneer, 
gligor, castelluccia2009difficulty} requires no hardware support. These techniques typically 
use precise timing measurements for round-trip time and attestation result computation.
Due to some strong assumptions, such RA techniques are only useful for one-hop attestation.

Hardware-based RA \cite{tan2011tpm,noorman2013sancus,kil2009remote,MQY10} relies on hardware 
components, such as Trusted Platform Modules (TPMs) \cite{tpm}, ARM Trustzone \cite{trustzone}, 
or Intel \cite{sgx}. This type of RA provides stronger security guarantees and lower overhead 
than software-based RA. However, it depends on relatively sophisticated and costly 
hardware components which may not be present in low-end or even mid-range IoT devices. 

Hybrid RA \cite{smart,vrasedp,tytan,trustlite} supports RA via both software and hardware features.
It typically uses hardware components to provide a RoT or a secure execution environment 
for software components. 

\noindent \textbf{Other Authentication Schemes for IoT Devices:}
Several authentication methods for IoT devices have been proposed e.g. \cite{shi2017smart, dhillon2017lightweight, al2023lightweight, alshahrani2021secure, eman2023multi, almadhoun2018user}, based on a variety of features such as biometrics, physical unclonable functions, channel characteristics, one-time passwords (OTPs), blockchain etc. \cite{shi2017smart} develops a deep learning based user authentication scheme that utilizes
WiFi signals to capture unique human physiological and behavioral characteristics inherited from their daily activities. \cite{al2023lightweight} 
presents a two-factor authentication protocol for IoT-enabled healthcare ecosystems 
using biometrics and post-quantum cryptography. \cite{alshahrani2021secure} proposes a 
lightweight and secure multi-factor device authentication protocol for IoT devices 
using configurable PUFs and channel-based parameters. \cite{eman2023multi} introduces 
a multi-device user authentication mechanism for IoT devices using OTPs and a 
novel device usage detection mechanism. \cite{almadhoun2018user} proposes a system that authenticates user access to IoT devices using blockchain-enabled fog nodes. The nodes connect to Ethereum smart contracts, which issue access tokens without requiring an intermediary or trusted third party.

%% file: content/08-conclusion.tex
\section{Conclusion}
This paper constructed \system, a secure multi-user access mechanism for a range of IoT devices. 
\system contends with hardware resource constraints of IoT devices and 
significant overhead associated with Kerberos. It involves a new component -- an IoT Server (\iotserver) 
-- a special Kerberos service responsible for managing access to IoT devices. 
We implemented an open-source prototype \cite{kesic-repo} of \system, which includes \iotserver, 
two IoT devices based on ARM Cortex-M33 equipped with TrustZone-M, and a client application. 
Its evaluation shows that a general device takes only 8.45ms to verify an IoT ticket, 
while a power-constrained device takes only 3.61ms.

\noindent {\bf Acknowledgements:}  We thank ICCCN’24 reviewers for constructive
feedback. This work was supported in part by funding
from NSF Award SATC-1956393, NSA Awards H98230-20-1-0345 and H98230-22-1-0308,
as well as a subcontract from Peraton Labs.